**Evaluation of the Gradient Boosting of Regression Trees Method on Estimating the Car Following Behavior**

**Sina Dabiri (Corresponding Author)**
Ph.D. Student
Charles Edward Via, Jr. Department of Civil and Environmental Engineering,
Virginia polytechnic institute and state university
750 Drillfield Drive, 301 Patton Hall, Blacksburg, VA 24061
Tel: (540) 231-6635; Fax: (540) 231-7532; Email: sina@vt.edu

**Montasir Abbas**
Associate Professor
Charles Edward Via, Jr. Department of Civil and Environmental Engineering,
Virginia polytechnic institute and state university
750 Drillfield Drive, 301-A Patton Hall, Blacksburg, VA 24061
Tel: (540) 231-9002; Fax: (540) 231-7532; Email: abbas@vt.edu

Word count:  6,074 words text + 5 tables/figures x 250 words (each) = 7,324 words

Submission Date: 07/31/2017

**Abstract**
Car-following models, as the essential part of traffic microscopic simulations, have been utilized to analyze and estimate longitudinal drivers' behavior since sixty years ago. The conventional car following models use mathematical formulas to replicate human behavior in the car-following phenomenon. Incapability of these approaches to capturing the complex interactions between vehicles calls for deploying advanced learning frameworks to consider the more detailed behavior of drivers. In this study, we apply the Gradient Boosting of Regression Tree (GBRT) algorithm to the vehicle trajectory data sets, which have been collected through the Next Generation Simulation program, so as to develop a new car-following model. First, the regularization parameters of the proposed method are tuned using the cross-validation technique and the sensitivity analysis. Afterward, the prediction performance of the GBRT is compared to the world-famous GHR model, when both models have been trained on the same data sets. The estimation results of the models on the unseen records indicate the superiority of the GBRT algorithm in capturing the motion characteristics of two successive vehicles.

## INTRODUCTION

The drivers' stochastic behavior renders more complexity, uncertainty, and non-linearity in transportation systems. Microscopic simulation architectures are key tools to replicate the stochastic nature of traffic conditions (1, 2) . These frameworks allow traffic engineers to evaluate the performance of the real-world traffic networks under various scenarios by virtually simulating the interactions between individual vehicles. This results in saving high expenditure on collecting field data and monitoring real transportation systems. In microscopic models, details of each vehicle's motion characteristics such as speed, acceleration/deceleration, time and space headway are described with the help of behavioral models including gap-acceptance, speed adaption, lane-changing, ramp metering, overtakes, and car-following models (3). Lateral and longitudinal interactions between vehicles are the two principal components for understanding dynamic behavior of travelers in traffic microscopic simulators (4). Lateral behaviors, described by lane changing models, comprise the decision for changing a lane, choice of a target lane, and gap acceptance (5). Car-following models reproduce longitudinal behavior of vehicles, in which movement features of the following vehicle are restricted to its leading vehicle in the same lane. The focus of this study is on car-following models as the principal type of microscopic simulation models.

A growing trend towards developing car-following models has been observed since sixty years ago (6). Many of classical car following models have been constructed according to traffic flow theories and explicit assumptions. Examples of conventional car following models range from Gazis–Herman–Rothery (GHR) models (7), safety distance or collision avoidance models (8), desired measures (9, 10), optimal velocity model (11), and linear models (12) to cellular automaton models (13), and psychophysical or action points models (14). The relative distance and speed between following and leading vehicles, the speed of the following vehicle as well as the driver's perception/reaction time are the standard inputs while the following vehicle's acceleration/deceleration or speed are the main outputs in the car-following models. On the other hand, the rapid growth of data-driven approaches in addition to availability of field data have led to proliferation of deploying data mining techniques directly in transportation fields (15, 16). In contrast to classical models, closed-form mathematical functions, calibrating, and manual errors are not concerns in data-driven techniques (17). Capability to extracting more details of drivers' response as well as flexibility for involving additional information (e.g., environmental effects and roadway geometry) are other benefits of using data-driven techniques in car-following models.

This study begins by reviewing car-following literature with focus on models developed with the aid of data-driven approaches. It will then present the Gradient Boosting of Regression Trees (GBRT) algorithm and the GHR model in the Methodology section. Trajectory data sets and preprocessing procedures are described in the Data Description section. In the Results and Discussion section, the optimal parameters of the proposed model are determined using the sensitivity analysis and the cross-validation technique. The prediction performance of the proposed model is also compared with the GHR model in this section. Finally, the conclusion is made in the Conclusion section.

## LITERATURE REVIEW

As mentioned in the previous section, car following models can be categorized into two main groups: conventional and data-driven models (4). A systematic review on the classical car-following models, their calibration, and evaluation have been reported by Brackstone et al. (6). Studies by Olstam et al. (3) and Panwai et al. (18) enlightened and compared the performance of classical car-following models in commercial traffic microsimulation packages including

AIMSUM with the safety distance model, MITSIM with the GHR model, VISSIM and Paramics with the psycho-physical models. In (19), a group of conventional car-following models such as GHR, Gipps, Cellular Automaton, S-K, Intelligent Driver, and Wiedemann models has been assessed in terms of the fundamental diagrams and the vehicle trajectory data sets. A qualitative study by Saifuzzaman (20) described extension and improved versions of the classical car following models from both engineering and human factors points of view. In accordance with the scope of the current study and existence of holistic survey papers on the classical car following models, we investigate only the cutting-edge car following models developed by data mining techniques.

Fuzzy logic, Artificial Neural Network (ANN), and combinations of the two approaches, (e.g., locally linear neural-fuzzy models (21)), have widely been utilized to simulate future states of a vehicle in a car-following process. A number of studies have also tested the performance of other machine learning algorithms, including k-nearest neighbor (17), locally weighted regression (22), and support vector regression (23) for the better understanding of drives' dynamic behaviors.

The reaction of a driver to a given stimulus is stemming from a set of ambiguous and vague driving rules rather than a deterministic relationship, which is typically enforced in the classic car-following models. To overcome such an issue, Kikuchi et al. (24), for the first time, presented a fuzzy inference system to simulate the driver's action as a result of a fuzzy reasoning process determined by natural-language-based driving rules. In their study, the relative distance and speed between the two vehicles as well as acceleration/deceleration of the leading vehicle were fed as inputs into the GHR model. Each of the inputs was grouped into six fuzzy sets. A triangular membership function was assigned to each fuzzy set. Specifying a set of rules generates a set of outputs in fuzzy numbers. Then, averaging weighted outputs over the time, as the final output, leads to estimating the following-vehicle movement relative to the leading vehicle. More work in this context can be found in (25, 26). The main challenge in developing such models was the determination of fuzzy sets and the corresponding membership functions, which are basic and essential concepts in the fuzzy logic.

The stochastic and non-linearity nature of human behavior and traffic streams call for more sophisticated models. Artificial Neural Network (ANN) has received the most attention in modeling the car-following behavior in comparison with other data-driven approaches due to its remarkable ability in deriving meaningful concepts from complicated and incomplete data sets. Hongfei et al. (27) introduced the application of ANNs in the car-following models for the first time. A two-hidden-layer back propagation ANN was applied to develop a car-following model, where the training and test data sets were collected by means of a five-wheel system in an open roadway condition. Moreover, the performance of other types of ANN architectures including Fuzzy Adaptive Resonance Theory (ART) neural networks, Radial basis function networks, and swarm algorithm have been investigated. Yet the results implied the superiority of the regular neural network compared to almost all other ANN training algorithms (28, 29). As ANNs require no pre-settings for considering the various combination of input variables or adding new features, the contribution of other factors, including the motion features of the preceding vehicle in front of the leading vehicle, can also be examined using ANNs (30). Unlike many studies that put the reaction time of the driver-vehicle unit as a constant value, a number of modified ANNs have been built to predict the car-following behavior based on the instantaneous reaction delay (31, 32). Several studies have attempted to fuse ANNs with other approaches such as fuzzy logic and local linear regression models (4, 33)

Notwithstanding that several data-driven approaches for modeling the car-following behavior have been postulated in literature, lack of examining ensemble algorithms in this context has encouraged us to evaluate the capability of this class of machine learning algorithms. Since all variables in the context of car following models are continuous, regression models are more reliable algorithms to build the relationship between the variables. Accordingly, the GBRT, as an effective type of ensemble algorithms, is deployed that takes advantages of predictive power, robustness to the noise by combining a set of base learners, and minimizing the prediction error as learning proceeds.

## METHODOLOGY

### GHR Models

The general GHR model, proposed by Gazis, Herman, and Rothery at General Motors laboratory (7), is used as the basis of comparison for examining the performance of the GBRT-based car following model. All GHR models express a general idea that a driver responds to a stimulus according to the following relationship: Response = function (sensitivity, stimulus), which also known as the follow-the-leader theory of traffic. The response is always considered as the acceleration/deceleration of the following vehicle, which is a factor that can be easily controlled by drivers through accelerator or brake pedals. According to the follow-the-leader theory, it has been shown that there is a high correlation between the following driver's response and the relative speed difference of the follower and leader vehicles. This results in considering the relative speed as the stimulus in all versions of the GHR models (7). Therefore, the variations among GHR models is contingent on how to formulate the sensitivity term. While a variety of functional forms were proposed and evaluated as the sensitivity term, the general GHR model was introduced to cover all other types of GHR models:

$$acc_{n+1}(t+\tau) = \alpha v_{n+1}^m(t+\tau) \frac{\Delta v_{n,n+1}(t)}{\Delta x_{n,n+1}^l(t)} \tag{1}$$

Denoting the current time as $t$, $acc_{n+1}(t+\tau)$ and $v_{n+1}(t+\tau)$ are the acceleration/deceleration and speed of the following vehicle at the instant $(t+\tau)$, respectively. $\tau$ is the time lag of driver's response to observed stimulus predictors. $n+1$ and $n$ are subscripts for the following and leading vehicles. $\Delta v_{n,n+1}(t)$ and $\Delta x_{n,n+1}(t)$ respectively represent the relative speed and the relative distance between the two vehicles at the time instant $t$. The model contains three parameters $\alpha$, $m$, and $l$ that need to be calibrated through the problem data set. Although a vast majority of studies have investigated the GHR model in an attempt to determine the best combination of the parameters, we define the optimal values by solving the optimization problem formulated in the equations (2) and (3). Optimizing the parameters for each data set separately, instead of using pre-defined values, leads to capturing the car-following behavior that is compatible with the traffic conditions exist in the dataset.

$$\min_{\alpha,m,l} \sum_{t=1}^{T} \left[ \alpha v_{n+1}^m(t+\tau) \frac{\Delta v_{n,n+1}(t)}{\Delta x_{n,n+1}^l(t)} - acc_{n+1}(t+\tau) \right]^2 \tag{2}$$

$$\text{S.t}: \ \mathbf{LB} \le \alpha,m,l \le \mathbf{UP} \tag{3}$$

where $T$ is the number of training instances. The equation (2) measures the sum of the squares of the errors between the estimated and true (observed) values of acceleration/deceleration. The equation (3) constrains the model parameters between the lower bound vector, $\mathbf{LB} = [0, 0, 0]$, and upper bound vector, $\mathbf{LB} = [3, 3, 3]$. The vectors have designed in such a way that cover all values used in the literature of GHR models, which have been summarized in the reference (6).

**Gradient Boosting of Regression Trees**

In this study, the response and explanatory variables of the proposing car-following model are considered similar to the GHR model. Accordingly, the acceleration/deceleration of the following vehicle at the time instant $(t + \tau)$, $acc_{n+1}(t + \tau)$, is the model's response variable predicted as a function of the explanatory variables, $[v_{n+1}(t + \tau), \Delta v_{n+1}(t), \Delta x_{n+1}(t)]$. Since all variables, particularly the response variable, are continuous, regression models are suitable approaches to predict car-following behavior. In this paper, we assess the performance of an ensemble regression algorithm in estimating car-following models.

The core idea in ensemble methods is to combine multiple base learners, which are usually called weak learners, in order to improve the accuracy and robustness of the final model. Two major questions need to be answered in ensembles methods: what type of weak learner should be used for the training process? How to train and fuse weak learners to yield the final model? Accordingly, ensemble algorithms fall into the two main families: averaging and boosting methods. Each family follow a distinct procedure towards training and combining individual base learners (34). In the averaging methods, base learners are first built independently on randomly selected training instances, and then are averaged to generate the final model. Examples of this category are bagging and random forest methods. Naïve Bayes classifiers, decision trees, and neural network are typically utilized as the machine learning techniques for training weak learners. In the boosting methods, weak learners are trained iteratively and in a stage-wise fashion so as to find a model that reduces the bias and variance of prediction. In each iteration, a strength weight is assigned to the weak learner that implies its prediction error rate; meanwhile, each training instance is reweighted by how incorrectly it was classified. The method of weighting training data and base learners distinguishes between boosting algorithms such as the adaptive boosting, and the gradient boosting. The final model is the sum of the weighted weak learners (34).

In this study, we envision to utilize the Gradient Boosting of Regression Trees (GBRT), which was initially developed by Friedman (35), for estimating the car-following model. The key goal in the GBRT algorithm is to, at each step, fit a regression tree to the difference between the observed response and the aggregated prediction of all learners grown previously. As indicated by the name of the algorithm, the decision trees algorithm with a fixed size is chosen as the weak leaner.

Suppose the training data set contain $N$ samples of $(y_i, \mathbf{x}_i)$, in which $y_i$ and $\mathbf{x}_i$ are the response variable and explanatory vector for the sample $i$, respectively. The goal is to learn a model $F^*(\mathbf{x})$, the best approximation of the real function $F(\mathbf{x})$ that maps $\mathbf{x}$ to y, while the expected value of the loss function $L(y, F(\mathbf{x}))$ is minimized as shown in the equation (4):

$$F^* = \underset{F}{\arg\min} \, E_{y,\mathbf{x}} L(y, F(\mathbf{x})) = \underset{F}{\arg\min} \, E_{y,\mathbf{x}} [(y - F(\mathbf{x})]^2 \tag{4}$$

Notwithstanding that many forms of loss functions exist in the literature, the squared-error loss function is chosen in the equation (4) due to the superior computational properties of the least-square algorithm for minimizing the square-error loss function (35). The GBRT algorithm seeks for finding the function $F(\mathbf{x})$ in the form of the sum of the weighted weak learners, $h(\mathbf{x}; \mathbf{a}_m)$:

$$F(\mathbf{x}; \{\beta_m, \mathbf{a}_m\}_1^M) = \sum_{m=1}^{M} \beta_m h(\mathbf{x}; \mathbf{a}_m) \tag{5}$$

where $M$ is the number of weak learners, and $\beta$ is the corresponding weight for each learner. $h(\mathbf{x}; \mathbf{a}_m)$ is a small regression tree defined by a finite set of parameters $\mathbf{a}_m$, including splitting variables, split locations, and leaf nodes. Analogous to other boosting algorithms, the GBRT builds an additive model in the form of the equation (6), when $F_m(\mathbf{x})$ in the last iteration equals to $F(\mathbf{x})$ in the equation (5):

$$F_m(\mathbf{x}) = F_{m-1}(\mathbf{x}) + \beta_m h(\mathbf{x}; \mathbf{a}_m) \tag{6}$$

Thus, at each step $m$, the estimator $h(\mathbf{x}; \mathbf{a}_m)$ is added to the imperfect model $F_{m-1}(\mathbf{x})$ in the previous step to build a better additive model in a forward stage-wise fashion presented in the equation (6). The decision tree $h(\mathbf{x}; \mathbf{a}_m)$ and its associated weight $\beta_m$ are trained to minimize the loss function of observed responses and $F_m(\mathbf{x})$ as shown in the equation (7):

$$[h(\mathbf{x}; \mathbf{a}_m), \beta_m] = \arg\min_{h, \beta} \sum_{i=1}^{N} L(y_i, F_{m-1}(\mathbf{x}_i) + \beta h(\mathbf{x}_i; \mathbf{a}_m)) \tag{7}$$

where the loss functions in the equations (4) and (7) are the same. As the equation (7) is a hard optimization problem, one way to resolve the problem easier is to apply the gradient descent algorithm. Considering the equation (6) and given the estimator $F_{m-1}(\mathbf{x})$, the function $\beta_m h(\mathbf{x}; \mathbf{a}_m)$ can be viewed as the best greedy step towards the best approximation of $F_m(\mathbf{x})$. In order to apply the gradient descent algorithm to the equations (6) and (7), $h(\mathbf{x}; \mathbf{a}_m)$ is trained using the training set $\{(\mathbf{x}_i, g_m(\mathbf{x}_i)\}_i^N$, where $g_m(\mathbf{x}_i)$ is the negative gradient of loss function computed by the eqaution (8):

$$g_m(\mathbf{x}_i) = -\left[ \frac{\partial L(y_i, F(\mathbf{x}_i))}{\partial F(\mathbf{x}_i)} \right]_{F(x) = F_{m-1}(\mathbf{x})} = [y_i - F_{m-1}(\mathbf{x}_i)] \tag{8}$$

where the right-handed side in the equation (8) is computed by substituting the squared-error loss function. Hence, at each step $m$, the regression tree $h(\mathbf{x}; \mathbf{a}_m)$ is fitted to the difference between the observed response and the aggregated prediction of all learners grown so far, (i.e., $h(\mathbf{x}; \mathbf{a}_m)$ is trained using the training set $\{(\mathbf{x}_i, g_m(\mathbf{x}_i)\}_{i=1}^N$).

The step size, $\beta_m$, is calculated by solving the one-dimensional optimization problem using the equation (9):

$$\beta_m = \arg\min_{\beta} \sum_{i=1}^{N} L(y_i, F_{m-1}(\mathbf{x}_i) + \beta h(\mathbf{x}_i; \mathbf{a}_m)) \tag{9}$$

It should be noted that solving the equation (9) can be thought of as finding the optimal step size in the gradient descent algorithm.

The GBRT algorithm steps are summarized as follows:

1. Initialize the approximation function $F(\mathbf{x})$:

$$F_0(\mathbf{x}) = \arg\min_{\beta} \sum_{i=1}^{N} L(y_i, \beta) \text{ or } F_0(\mathbf{x}) = \frac{1}{N}\sum_{i=1}^{N} y_i$$

2. For $m = 1$ to $M$ do:

- Calculate the pseudo-responses: $g_m(\mathbf{x}_i) = \left[ y_i - F_{m-1}(\mathbf{x}_i) \right]$ for $i = 1, \ldots, N$

- Fit the regression tree $h(\mathbf{x}; \mathbf{a}_m)$ using the training set $\{(\mathbf{x}_i, g_m(\mathbf{x}_i)\}_{i=1}^{N}$

- Calculate the step size $\beta_m$ using the line search: $\beta_m = \arg\min_{\beta} \sum_{i=1}^{N} L(y_i, F_{m-1}(\mathbf{x}_i) + \beta h(\mathbf{x}_i; \mathbf{a}_m))$

- Update the model: $F_m(\mathbf{x}) = F_{m-1}(\mathbf{x}) + \beta_m h(\mathbf{x}; \mathbf{a}_m)$

3. End the algorithm. $F_M(\mathbf{x}) = F^*(\mathbf{x})$ is the final output of the algorithm.

**Regularization Parameters**

Regularization is a process used in an attempt to prevent the overfitting problem in statistical models by explicitly controlling the model complexity and constraining the fitting procedure (35). Overfitting occurs when the model is extremely complex with too many parameters. Large number of parameters engenders the learning process to be performed through memorization of the training data set rather than understanding the concepts and attributes inside the data. An over-fitted model perfectly predicts only on training examples due to the memorization while has a poor predictive performance on the unseen data. For the GBRT model, adding too many weak learners with over-complex trees as well as the high speed of learning are the main sources of overfitting, which call for introducing regularization parameters to control the issue.

The value of $M$, the number of weak learners, is one of the regularization parameters that regulates the expected loss reduction over the training data. Moreover, handling the speed of learning through a shrinkage process has been found to be an efficient approach for avoiding complexity and overfitting, particularly in additive models (35). The shrinkage strategy can be obtained by incorporating the learning rate v, which scales the contribution of each weak learner, as shown in the equation (10). Thus, the additive model in the second step of the algorithm is replaced by:

$$F_m(\mathbf{x}) = F_{m-1}(\mathbf{x}) + v.\beta_m.h(\mathbf{x}; \mathbf{a}_m) \qquad 0 < v \leq 1 \qquad (10)$$

The parameters $M$ and $v$ interact strongly with each other. Thus, the value of each affects the other. Accordingly, a sensitive analysis that considers both parameters simultaneously needs to be performed for determining their optimal values. Model evaluation techniques such as cross-validation is being deployed to compute and compare the generalization error under each combination of $M$ and $v$.

The size of regression trees, denoted as $D$, is another regularization parameter that controls the overfitting problem. The tree depth is defined by tuning parameters like the maximal number of decision splits and/or the minimum number of observations required to be at a leaf node. The decision tree is usually over-fitted on the dataset with a few samples yet a large number of features. Fortunately, the current dataset contains a large number of samples with only three features, which reduces the possibility of generating over-fitted trees. Even though the size of trees has a slight impact on the overfitting problem due to the mentioned characteristics of the car-following dataset, we compute its optimal value by performing a sensitivity analysis. So we not only obtain the highest decrease in the test error but ensure that the important structures of the ultimate model are captured (36).

It is worth mentioning that, in this paper, the driver's reaction time ($\tau$) is not an explanatory variable in both GHR and GBRT models; instead, it is used as a constant value in the models. Indeed, its value affects the order of predictor variables with their corresponding response variable. For example, the triple $\left[v_{n+1}(t+\tau), \Delta v_{n+1}(t), \Delta x_{n+1}(t)\right]$ is used as input variables for predicting the following car acceleration at the time instant $(t+\tau)$, $acc_{n+1}(t+\tau)$. As a consequence, each data instance is defined as $\left\{acc_{n+1}(t+\tau), \left[v_{n+1}(t+\tau), \Delta v_{n+1}(t), \Delta x_{n+1}(t)\right]\right\}$, which is contingent on the value of $\tau$. Considering the reaction-time range between 0.1-3.0 seconds as the interval that covers all ranges offered by a variety of studies (22), the optimal value of $\tau$ is determined through a sensitivity analysis for both models. It is worth mentioning that optimal value of $\tau$ might be different for each model.

## DATA DESCRIPTION

The proposed car following model is developed and tested based on vehicle trajectory data sets collected by Federal Highway Administration (FHWA) under the Next Generation Simulation (NGSIM) program. The NGSIM program aimed to collect high-quality data on vehicles' motion parameters over a time window for which drivers' behaviors and their interactions with traffic objects are inferred through the microscopic modeling. The data in this study are selected from the northbound traffic on I-80 in Emeryville, California recorded on April 2005. NGSIM I-80 data set contains three subsets recorded in three time periods: 4:00 to 4:15 pm, 5:00 to 5:15 pm, and 5:15 to 5:30 pm. Deploying seven video cameras installed on a tall building, the trajectory data were recorded at the one-tenth-of-second resolution. A variety of vehicle trajectory features have been extracted and provided in each data set including acceleration, speed, location coordinates, ID of the preceding and the following vehicle of the subject vehicle, type and length of the vehicle, space and time headway with respect to the leading vehicle.

However, the NGSIM data include measurement errors in computing location coordinates, which in turn produce the unrealistic large speed and acceleration values. Raw data have also been affected by high-to-medium-frequency disturbances due to speed transitions, which needs to be modified. Several studies in literature have proposed techniques to rectify errors in the raw data before being used in traffic flow theory problems (37, 38). In this study, the proposed multi-step filtering procedure in the reference (37) is applied to the NGSIM data to create reconstructed reliable trajectory data. First, the data points that their absolute acceleration/deceleration exceeds a certain threshold value, which is 3 m/s$^2$ within this scope, are detected as unrealistic values. Using a filtering technique such as natural cubic spline interpolation, new longitudinal locations of outliers are interpolated between a sequence of non-outlier points before and after the outliers. The outlier speed and acceleration are then calculated from the new distance locations through the basic equations of motion. Secondly, a low-pass filter is utilized to remove the random error component from speed and acceleration profiles. Further details on the vehicle-trajectory-reconstruction procedure can be found in the reference (37).

In this study, the total of six data series, two from each subset, are opted for the analysis. In each of three subsets, two pairs of following and leading vehicles that have the highest number of data points were selected. The follower vehicle in one pair is an Auto, and in the other one is a Truck vehicle. Each pair builds one data series that ends up with the total six data series, named Auto-1, Truck-1, Auto-2, Truck-2, Auto-3, and Truck-3. Summary statistics of the acceleration, speed, and relative distance of the following vehicle in the original data series are provided in Table 1. Using the above-mentioned filtering procedure, the reconstructed data of speed and acceleration profiles for Auto-2 data series, as the data series with the maximum number of samples, are plotted against the original NGSIM data in Figure 1. The proposed car-following algorithm and the corresponding analysis are applied to all data series.

**Table 1  Basic Statistics for Acceleration, Speed, and Relative Distance of All Data Series**

| Data Series | No. of samples | Statistics for acceleration (m/s²) | | | Statistics for speed (m/s) | | | Statistics for relative distance (m) | | |
|---|---|---|---|---|---|---|---|---|---|---|
| | | min | max | mean | min | max | mean | min | max | mean |
| Auto-1 | 845 | -3.41 | 3.41 | -0.01 | 0.57 | 9.16 | 5.26 | 8.30 | 28.22 | 14.39 |
| Truck-1 | 473 | -3.41 | 3.41 | 0.03 | 4.56 | 9.23 | 6.85 | 3.41 | 38.24 | 25.76 |
| Auto-2 | 1444 | -3.41 | 3.41 | 0.02 | 0.00 | 3.97 | 1.31 | 6.43 | 13.62 | 9.93 |
| Truck-2 | 1130 | -3.41 | 3.41 | 0.00 | 1.36 | 6.71 | 3.42 | 25.08 | 41.58 | 32.91 |
| Auto-3 | 1412 | -3.41 | 3.41 | 0.00 | 0.00 | 3.07 | 1.80 | 5.68 | 14.70 | 10.08 |
| Truck-3 | 1415 | -3.41 | 3.41 | 0.04 | 0.67 | 7.25 | 2.74 | 6.12 | 15.14 | 10.16 |

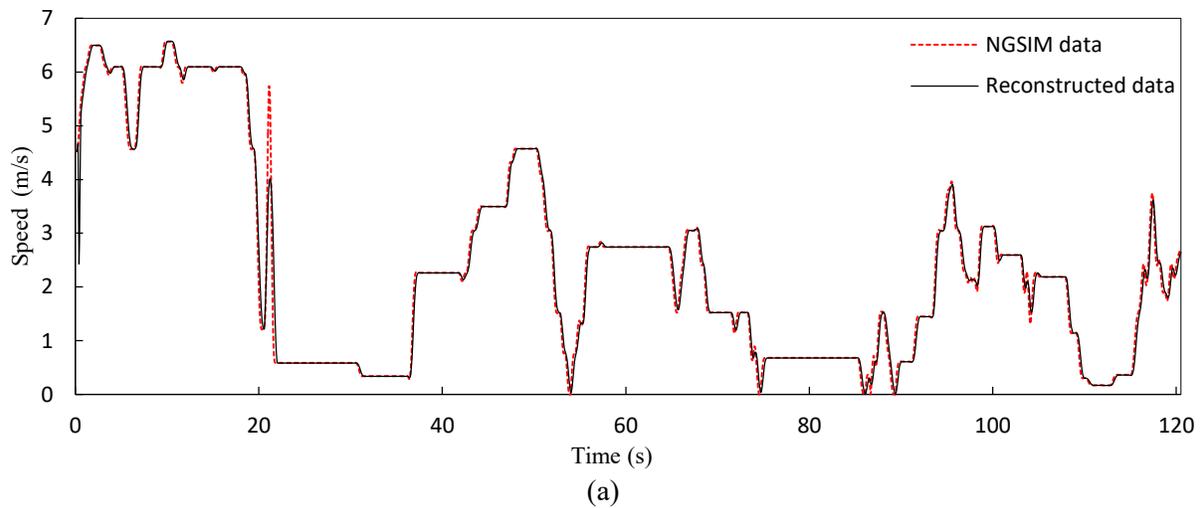

(a)

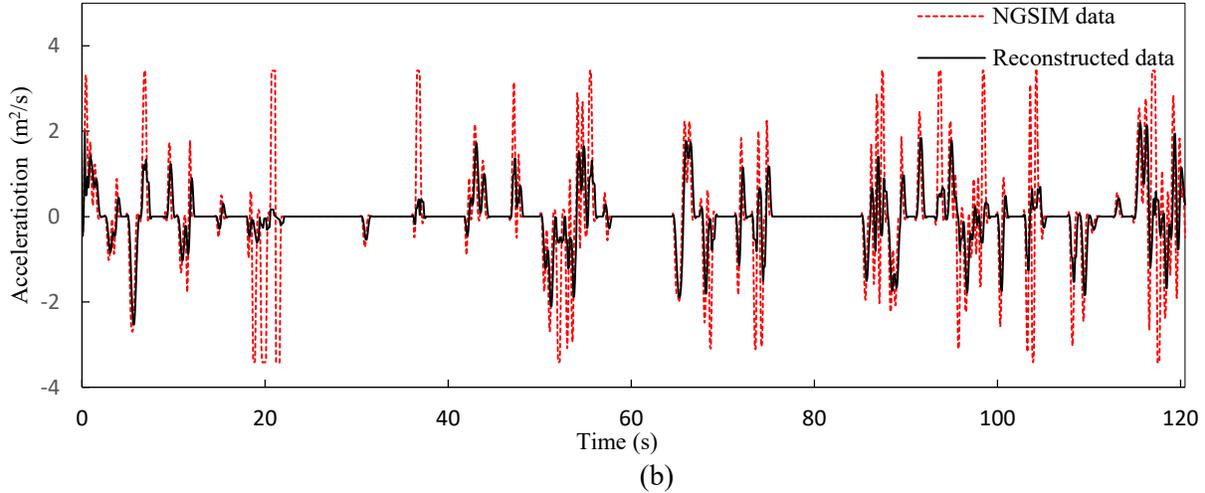

**FIGURE 1** **Comparison between the NGSIM and reconstructed data of speed and acceleration profiles for the Auto-2 data series: (a) Speed profile and (b) Acceleration profile.**

## RESULTS AND DISCUSSION

The model evaluation process for the proposed GBRT-based car-following model is implemented in two steps. First, the regularization parameters ($v$, $M$, $D$, $\tau$) are optimized using the 5-fold cross-validation to ensure a good fit of the data. In the second stage, using the optimal values found in the first stage, the performance of the GBRT model is compared with the optimum GHR model that has been optimized based on the equations (2) and (3). It should be noted the optimal reaction time is also determined for the GHR model to have a fair comparison. As mentioned in the methodology section, the data instances that are fed into the both algorithms vary according to the reaction time ($\tau$) value. Such a time dependency between consequent data instances hinders to split them randomly into the training and test sets. Consequently, in this study, the first 80% of each data series is selected as the training data set while holding out the remaining 20% of the data as the unseen records to be used only in the second step of the model evaluation process. Both steps are consecutively implemented on each of the six data series separately and their results are reported.

### First step: regularization parameters

In the sensitivity analysis for computing the regularization parameters, the 5-fold cross validation is utilized to obtain the average prediction error. Due to the inherent time dependency between the explanatory and response variables of the consequent data samples, unlike the general cross-validation, each fold is not created randomly. While the test set is still held out for the second step of the evaluation, the training set is subdivided into five subsets in such a way that the first 20% of the training portion of each data series constitutes the first fold, the second 20% forms the second fold and so forth.

A matrix with different combinations of $v$ and $M$ is designed for simultaneously optimizing the learning rate and the number of weak learners. The vectors $\mathbf{V} = [0.1, 0.3, 0.5, 0.8, 1]$ and $\mathbf{M} = [1:1:9, 10:10:90, 100:100:1000]$ constitute the rows (learning rate) and columns (the number of weak learners) of the matrix, respectively. For each element of the designed VM matrix, (i.e., a pair of ($v$, $M$)), the 5-fold cross validation is applied to the training set to measure the effectiveness of the model. At each iteration, the GBRT is fitted to the four folds and validated on the remaining fold by computing the Mean Square Error (MSE) as the performance metric. The final performance measure pertaining to each pair $(v, M)$ is the average of the MSE values computed in the loop.

Figure 2 illustrates how the GBRT prediction error is influenced by various combinations of $(v, M)$ for the data series Auto-2 and Truck-2, as the data sets with the highest number of samples. Each line represents the variation of the average MSE for a particular value of $v$ over the full range of the vector **M**. Initially, after a fluctuation for the values of $M$ less than 10, the average MSE increases gradually when the number of weak learners is also rising. However, the rate of average MSE levels out after $M$ goes up to the hundreds of weak learners. Intuitively, it was expected that the prediction performance is enhanced by raising the number of base learners. Nonetheless, the results indicate that augmenting the $M$ render the model over-complex and over-fitted, which in turn leads to deteriorating the accuracy of the GBRT model. Thus, the lower values of $M$ are more efficient in generalizing the GBRT model to unseen instances. According to the sensitivity analysis of $(v, M)$ for all data series, $M = 8$ is chosen as the optimum value for the further analysis. Furthermore, as shown in the Figure 2, the smaller values of the learning rate give the more improved predictive model. Although setting the learning rate to values lower than 0.1 will almost always ameliorate the prediction performance, we choose v = 0.1 as the optimal value to maintain the computational cost at a reasonable level (39). It is worth mentioning that the similar trends in Figure 2 have been obtained for other data series as well.

Figures (3a) and (3b) delineate how the accuracy of the GBRT model is influenced by changing the size of the decision trees ($D$) and the reaction time ($\tau$), respectively. The average MSE is computed for each value of $D$ and $\tau$ with the analogous approach to the $v$ and $M$. As depicted in Figure (3a), the average MSE for almost all data series declines slightly by setting $D$ equals to 3. However, the variation in MSE remains constant for values of $D$ greater than 3. Accordingly, the $D = 3$ is selected as the optimum value of the decision tree depth. Such a plateau stems from that fact that the car-following model considers only a few features. As a consequence, the possibility of facing the overfitting problem decreases even if the decision trees become over-complex by increasing the value of $D$. Furthermore, as depicted in Figure (3b), the GBRT prediction performance remained approximately constant over the reaction time interval although the lower values of the reaction time work better for some of the data series such as Auto-1 and Auto-2. It is clear that the optimal value of the reaction time is an essential characteristic of the driver in a given data set rather than the learning algorithm, which may be distinct for other data sets. Since the scope of this study is to tune the GBRT parameters instead of determining a reliable reaction time interval for all possible data sets, we use the exact optimal value of the reaction time obtained for each data series. In summary, as mentioned in the Methodology section, the most effective regularization parameters for the GBRT-based car following models are $(v, M)$ while the $(D, \tau)$ have not a significant impact on controlling the degree of fit to the data.

**Step 2: Comparison between GBRT and GHR models**

We, first, optimize the parameters of the GBRT and GHR models for establishing a fair comparison between the two predictive models, as one of the salient objectives in this study. Then, for having a robust comparison, the fitted models are examined on the test samples of each data series, which have not been deployed in the training process. For the test portion of each data series, the acceleration of the following vehicle is estimated with both techniques and the corresponding MSEs are reported as shown in Figure 4.

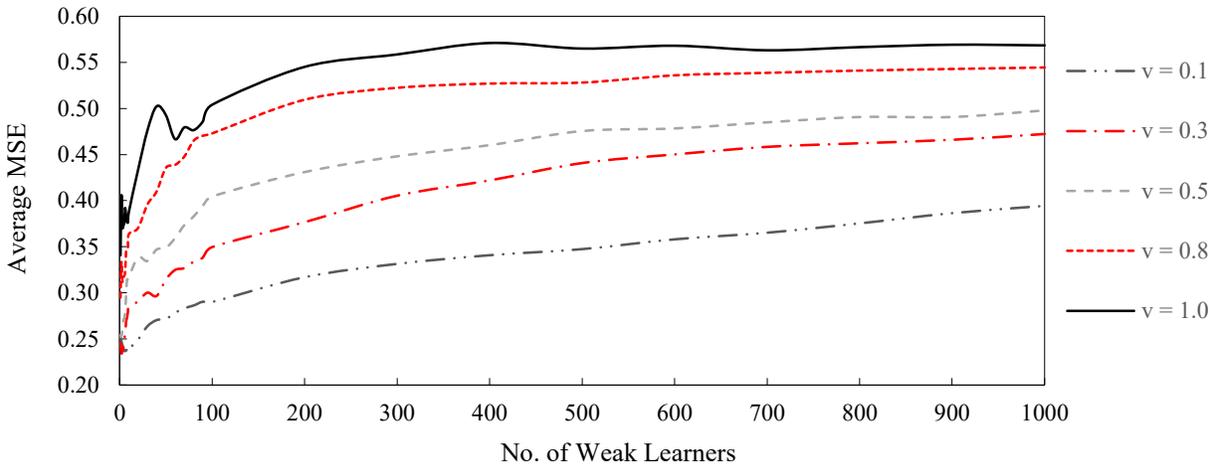

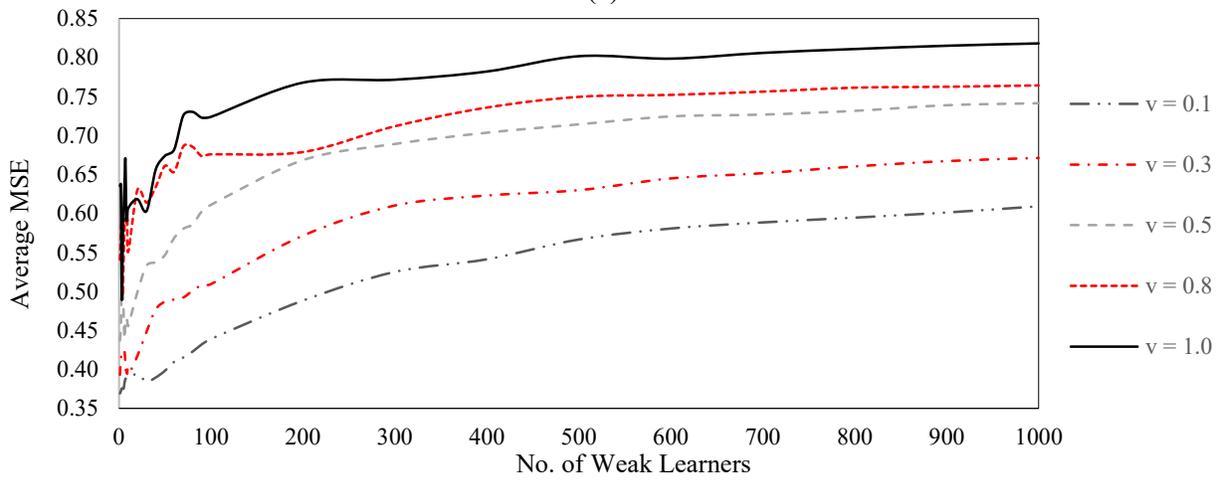

**FIGURE 2  Average MSE for the GBRT model upon various combinations of the number of weak learners (*M*) and the learning rate (*v*) for two data series: (a) Auto-2, (b) Truck-2.**

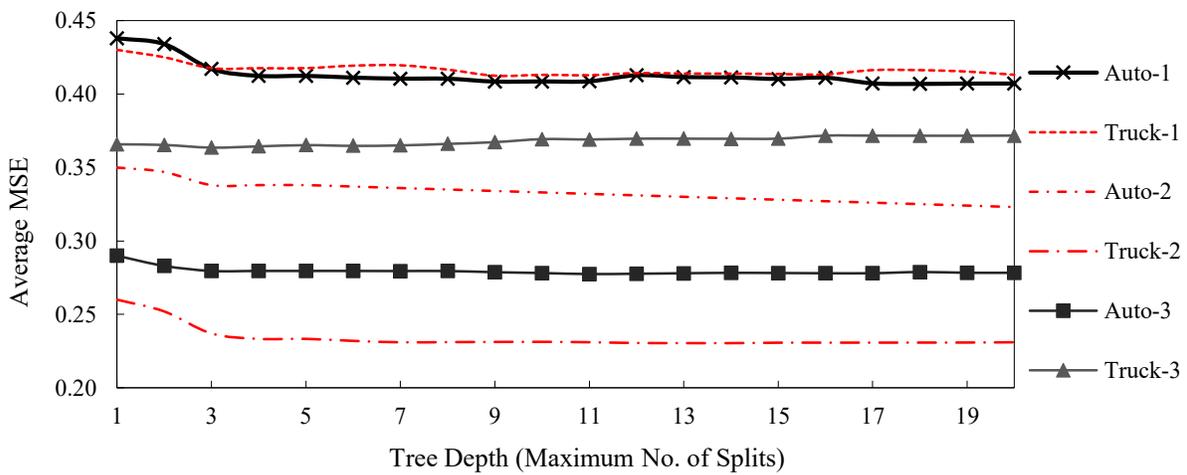

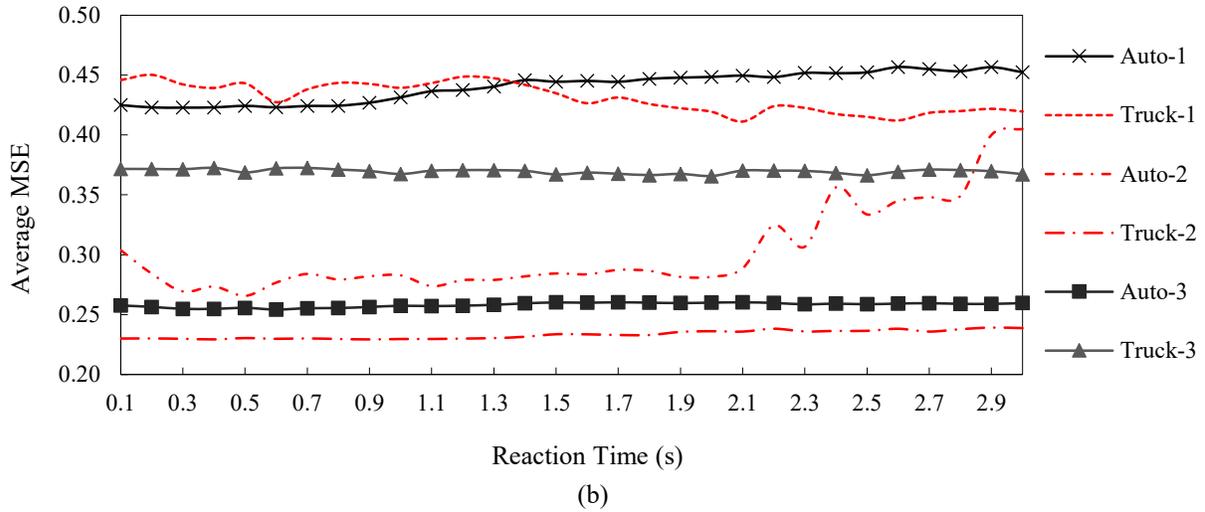

(b)

**FIGURE 3  Average MSE of the GBRT model upon different values of: (a) Tree Depth (Maximum Number of Splits), (b) Reaction Time (s).**

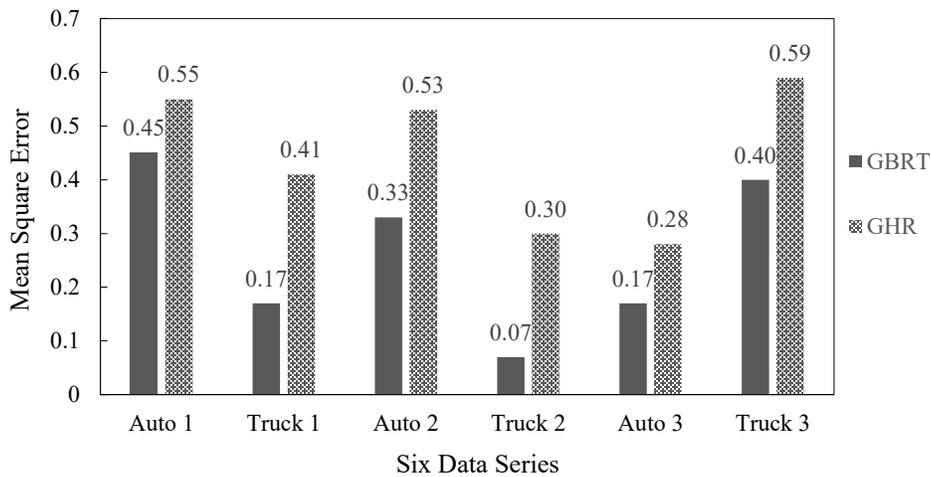

**FIGURE 4  Performance comparison between the GBRT and GHR Models based on MSE, as the good-of-fit measure, for six data series.**

The comparison between the two models proves the superiority of the GBRT model in all data series, regardless of whether the auto or truck is the follower vehicle. The primary reason of the GBRT preeminence originates from the fact that such models are unrestricted to grow according to the relationship among variables. However, constraining the interdependency between features of the car-following behavior to a pre-defined structure lessons the GHR power to capture more aspects of drivers' behavior.

**CONCLUSION**

With the emergence of advanced data-driven approaches and technologies for collating the massive vehicle trajectory data sets, the sophisticated car-following models can be developed so as to alleviate the shortcomings of the classical car following models such as restriction to mathematical equations and inability to incorporating additional effective factors. With the help of data mining techniques, there is no need to define the functionality form of the model. This is a practical benefit particularly in the complex traffic flow condition where building an appropriate model is almost impossible. Furthermore, data-driven approaches give room to researchers to investigate the effect of new factors on the car-following behavior, including other kinds of the microscopic and macroscopic parameters, weather conditions, leading vehicle types, time of day or week, road geometry, and the motion inputs related to the vehicles other than the leading vehicle.

In this study, the Gradient Boosting of Regression Tree algorithm was presented to reproduce the response action of the following vehicle in accordance with the motion characteristic of the leading vehicle. Data sets for the analysis were captured from the NGSIM program. Data pre-processing techniques were applied to the raw data to remove random error measurements and obtain a reliable data source. In order to ensure having a promising model, first, the regularization parameters of the model were determined. After that, the optimal GBRT and GHR models were trained on the same training set. The prediction results of applying both models to the unseen test data demonstrated the superiority of the proposed GBRT algorithm compared to the well-known GHR model.

As a future research direction, it would be interesting to simultaneously capture other drivers' stochastic behaviors. Using advanced machine learning algorithms, lateral and longitudinal behaviors of a vehicle can be simulated by integrating the lane-changing, car-following, and gap-acceptance concepts. Furthermore, outstanding frameworks for replacing the traditional models with the data-driven approaches in the commercial microsimulation packages such as VISSIM and CORSIM should be provided.

**REFERNCES**

1.      Dabiri S, Kompany K, Abbas M. Introducing a Cost-Effective Approach for Improving the Arterial Traffic Performance Operating Under the Semi-Actuated Coordinated Signal Control. 2018.

2.      Dabiri S, Abbas M, editors. Arterial traffic signal optimization using Particle Swarm Optimization in an integrated VISSIM-MATLAB simulation environment. Intelligent Transportation Systems (ITSC), 2016 IEEE 19th International Conference on; 2016: IEEE.

3.      Janson Olstam J, Tapani A. Comparison of Car-following models. Statens väg-och transportforskningsinstitut; 2004.

4.      Aghabayk K, Forouzideh N, Young W. Exploring a Local Linear Model Tree Approach to Car-Following. Computer-Aided Civil and Infrastructure Engineering. 2013;28(8):581-93.

5.      Ahmed KI. Modeling drivers' acceleration and lane changing behavior: Massachusetts Institute of Technology; 1999.

6.      Brackstone M, McDonald M. Car-following: a historical review. Transportation Research Part F: Traffic Psychology and Behaviour. 1999;2(4):181-96.

7.      Gazis DC, Herman R, Rothery RW. Nonlinear follow-the-leader models of traffic flow. Operations research. 1961;9(4):545-67.

8.      Gipps PG. A behavioural car-following model for computer simulation. Transportation Research Part B: Methodological. 1981;15(2):105-11.

9.      Koshi M, Kuwahara M, Akahane H. Capacity of sags and tunnels on Japanese motorways. ite Journal. 1992;62(5):17-22.

10.     Treiber M, Hennecke A, Helbing D. Congested traffic states in empirical observations and microscopic simulations. Physical review E. 2000;62(2):1805.

11.     Bando M, Hasebe K, Nakayama A, Shibata A, Sugiyama Y. Dynamical model of traffic congestion and numerical simulation. Physical review E. 1995;51(2):1035.

12.     Helly W. Simulation of bottlenecks in single-lane traffic flow. 1900.

13.     Nagel K, Schreckenberg M. A cellular automaton model for freeway traffic. Journal de physique I. 1992;2(12):2221-9.

14.     Wiedemann R, Reiter U. Microscopic traffic simulation: the simulation system MISSION, background and actual state. Project ICARUS (V1052) Final Report. 1992;2:1-53.

15.     Dabiri S, Heaslip K. Inferring transportation modes from GPS trajectories using a convolutional neural network. Transportation Research Part C: Emerging Technologies. 2018;86:360-71.

16.     Dabiri S, Heaslip K. Transport-domain applications of widely used data sources in the smart transportation: A survey. arXiv preprint arXiv:180310902. 2018.

17.     He Z, Zheng L, Guan W. A simple nonparametric car-following model driven by field data. Transportation Research Part B: Methodological. 2015;80:185-201.

18.     Panwai S, Dia H. Comparative evaluation of microscopic car-following behavior. IEEE Transactions on Intelligent Transportation Systems. 2005;6(3):314-25.

19.     Hamdar S. Driver behavior modeling.  Handbook of Intelligent Vehicles: Springer; 2012. p. 537-58.

20.     Saifuzzaman M, Zheng Z. Incorporating human-factors in car-following models: A review of recent developments and research needs. Transportation Research Part C: Emerging Technologies. 2014;48:379-403.

21.     Khodayari A, Ghaffari A, Kazemi R, Braunstingl R, editors. Modify car following model by human effects based on locally linear neuro fuzzy. Intelligent Vehicles Symposium (IV), 2011 IEEE; 2011: IEEE.

22.     Papathanasopoulou V, Antoniou C. Towards data-driven car-following models. Transportation Research Part C: Emerging Technologies. 2015;55:496-509.

23.     Wei D, Liu H. Analysis of asymmetric driving behavior using a self-learning approach. Transportation Research Part B: Methodological. 2013;47:1-14.

24.     Kikuchi S, Chakroborty P. Car-following model based on fuzzy inference system. Transportation Research Record. 1992:82-.

25.     Wu J, Brackstone M, McDonald M. Fuzzy sets and systems for a motorway microscopic simulation model. Fuzzy sets and systems. 2000;116(1):65-76.

26.     Gao Q, Hu S, Dong C, editors. The modeling and simulation of the car-following behavior based on fuzzy inference. Modelling, Simulation and Optimization, 2008 WMSO'08 International Workshop on; 2008: IEEE.

27.     Hongfei J, Zhicai J, Anning N, editors. Develop a car-following model using data collected by" five-wheel system". Intelligent Transportation Systems, 2003 Proceedings 2003 IEEE; 2003: IEEE.

28.     Panwai S, Dia H. Neural agent car-following models. IEEE Transactions on Intelligent Transportation Systems. 2007;8(1):60-70.

29.     Colombaroni C, Fusco G. Artificial neural network models for car following: experimental analysis and calibration issues. Journal of Intelligent Transportation Systems. 2014;18(1):5-16.

30.     TANAKA M, UNO N. Evaluation of Car-Following Input Variables and Development of Three-Vehicle Car-Following Models with Artificial Neural Networks. Journal of the Eastern Asia Society for Transportation Studies. 2015;11(0):1826-41.

31.     Khodayari A, Ghaffari A, Kazemi R, Braunstingl R. A modified car-following model based on a neural network model of the human driver effects. IEEE Transactions on Systems, Man, and Cybernetics-Part A: Systems and Humans. 2012;42(6):1440-9.

32.     Zheng J, Suzuki K, Fujita M. Car-following behavior with instantaneous driver–vehicle reaction delay: A neural-network-based methodology. Transportation Research Part C: Emerging Technologies. 2013;36:339-51.

33.     Chong L, Abbas MM, Flintsch AM, Higgs B. A rule-based neural network approach to model driver naturalistic behavior in traffic. Transportation Research Part C: Emerging Technologies. 2013;32:207-23.

34.     scikit-learn. Ensemble methods.  http://scikit-learn.org/stable/modules/ensemble.html Accessed by July, 2017.

35.     Friedman JH. Greedy function approximation: a gradient boosting machine. Annals of statistics. 2001:1189-232.

36.     Friedman J, Hastie T, Tibshirani R. The elements of statistical learning: Springer series in statistics Springer, Berlin; 2001.

37.     Montanino M, Punzo V. Making NGSIM data usable for studies on traffic flow theory: Multistep method for vehicle trajectory reconstruction. Transportation Research Record: Journal of the Transportation Research Board. 2013(2390):99-111.

38.     Punzo V, Borzacchiello MT, Ciuffo B. On the assessment of vehicle trajectory data accuracy and application to the Next Generation SIMulation (NGSIM) program data. Transportation Research Part C: Emerging Technologies. 2011;19(6):1243-62.

39.     Ridgeway G. Generalized Boosted Models: A guide to the gbm package. Update. 2007;1(1):2007.